\pdfoutput=1
\documentclass[11pt]{article}
\usepackage[preprint]{acl}
\usepackage{times}
\usepackage{latexsym}
\usepackage[T1]{fontenc}
\usepackage[utf8]{inputenc}
\usepackage{microtype}
\usepackage{booktabs}
\usepackage{amsmath,amssymb,amsfonts}
\usepackage{stmaryrd}
\usepackage{graphicx}

\title{TRACE: Transparent Retrieval for Abstract Concept Evaluation}

\author{Joseph Bingham \\ Dept. of Biology, Technion University \\ jbingham@campus.technion.ac.il}

\begin{document}
\maketitle

\begin{abstract}
Recent work reports that vision--language models (VLMs) struggle to establish and maintain stable reference in repeated reference games. Rather than ask which VLM does best, we ask a more basic question: do you need a large pretrained VLM for this at all? On grounding a single director utterance to one of twelve tangram silhouettes, we compare six off-the-shelf VLMs against a transparent baseline that uses \emph{no learned visual representation}: classical SIFT keypoint matching and a signal-quality index over retrieved images. On identical trials, the transparent baseline matches the strongest VLM (SigLIP-large) and significantly outperforms the other five, including every CLIP and OpenCLIP variant. The baseline additionally retrieves external images, so this is not a matched-information comparison; what it shows is that a learned \emph{visual} representation is not the bottleneck for this task: given retrieved images, a shape-appropriate classical similarity suffices. Along the way we find that abstract-grounding ability varies widely across VLMs (15--39\% top-1; chance 8.33\%, humans $\approx$77--80\%), so the weakness is model-specific rather than intrinsic to contrastive pretraining; on the 1{,}013-shape KiloGram benchmark the pattern generalizes for CLIP, with per-shape difficulty tracking human shape-nameability. The pipeline is a classical, inspectable alternative rather than a learned one. We close by sketching how an explicit, inspectable representation of listener-side pact state could carry this approach into interactive multi-turn reference, which we leave to future work. Code available in supplementary material.
\end{abstract}

\section{Introduction}
\label{sec:intro}

Repeated reference games have served psycholinguistics as a controlled paradigm for studying how conversational partners establish common ground for over four decades \citep{clark1986,brennan1996,hawkins2020}. In the paradigm, a director and matcher see the same set of abstract stimuli (typically tangram silhouettes) in different orders; the director describes one target per trial, and the matcher must identify it. Over rounds, pairs converge on stable, partner-specific names for each figure, a process called \emph{lexical entrainment} that produces \emph{conceptual pacts}.

Whether contemporary vision--language models exhibit anything like this convergence has been the subject of substantial recent empirical work. \citet{imai2025} introduce a four-metric suite for interactive grounding and show that leading proprietary VLMs diverge from human patterns on at least three of the four in self-play. \citet{zeng2026} report systematic failures of LVLMs to shorten references or manage pact state across turns in referential communication. Closest to the present work, \citet{zhao2025} run five leading VLMs (GPT-4o, GPT-5-mini, Claude-3.7-sonnet, Claude-sonnet-4, Gemini-2.5-flash) on the exact tangram corpus we study and find that while human--human dyads improve from 78\% to 96\% accuracy across six rounds, agent--agent dyads fail to exhibit comparable convention formation.

These failures, among others LM domain failures~\cite{hacohen2026taking}, motivate the question we address: on the subtask of grounding a single director utterance to the intended tangram, how well do off-the-shelf vision--language models do, and does the task require a learned visual representation at all? This is a diagnostic, not a matched benchmark: our aim is to ask what information a system needs to ground abstract reference, not to replace VLMs in practice. The six VLMs receive only the utterance and the twelve candidate images, while our baseline additionally retrieves external images and scores them with a classical metric, so the two answer different questions and consume different information by design. The finding is that a classical pipeline reaches the accuracy of the strongest VLM and exceeds the rest: competitive grounding on this task without a learned visual representation in the matching stage. Along the way we characterize how sharply grounding varies across VLMs and show the pattern generalizes beyond the twelve tangrams. We close by sketching, as future work, an inspectable representation of listener-side pact state for the interactive setting.

\paragraph{Contributions.}
\begin{itemize}
    \item \textbf{A transparent baseline that matches large VLMs.} On identical trials, TRACE, a classical retrieval-plus-similarity pipeline with \emph{no learned visual representation} (SIFT alignment, Universal Quality Index), matches the strongest off-the-shelf VLM (SigLIP-large) and significantly outperforms every CLIP and OpenCLIP variant (\S\ref{sec:main}, Table~\ref{tab:main}). It is a classical, inspectable alternative rather than a learned one. An in-retrieval similarity control (\S\ref{sec:pipeline}) shows that access to the same retrieved images does not by itself help a learned similarity (CLIP or SigLIP); a fuller decomposition of retrieval, external knowledge, and similarity is left to future ablations.
    \item \textbf{A diagnostic of VLM abstract-reference grounding.} Across six off-the-shelf models, top-1 accuracy ranges from 15\% (OpenCLIP) to 39\% (SigLIP), against chance 8.33\% and a human rate of $\approx$77--80\% (\S\ref{sec:clip}, \S\ref{sec:fair}). SigLIP substantially outperforms every CLIP variant, so weak grounding is model-specific, not a property of contrastive image--text pretraining; domain-anchoring prompts and orientation augmentation both \emph{reduce} accuracy, while scale helps within a family.
    \item \textbf{Generalization and a link to human nameability.} On the 1{,}013-shape KiloGram benchmark \citep{ji2022kilogram} the CLIP result generalizes (\S\ref{sec:kilogram}), and per-shape grounding difficulty correlates with Shape Naming Divergence (a human nameability norm), and does so more strongly for the more capable backbone: models find the shapes hard that people find hard to name.
\end{itemize}

\section{Related Work}
\label{sec:related}

\paragraph{Grounding failures in LVLMs.}
The most direct motivation for this paper is the recent line of work documenting that LVLMs form common ground unlike humans \citep{imai2025,zeng2026,zhao2025}: across grounding-metric suites, human/AI dyad comparisons, and the Stanford tangram corpus we use, these studies report systematic failures at reference shortening, pact management, and agent--agent convention formation.

\paragraph{Abstract visual reasoning and tangrams.}
Tangrams have long served as cognitive-science stimuli, and \citet{ji2022kilogram} scale them into KiloGram, a benchmark of over a thousand annotated tangram shapes with whole-shape and part-level descriptions and a per-shape nameability norm, Shape Naming Divergence (SND). They report that pretrained multimodal models show limited abstract reasoning on tangrams, improving substantially with fine-tuning and with part-level descriptions. Our CLIP result is consistent with theirs. What we add on the perceptual side is a diagnostic on the \emph{repeated-reference} corpus of \citet{hawkins2020} in particular, controls showing that prompting and orientation do not rescue CLIP (a complement to their fine-tuning result), and a generalization to the full KiloGram set that ties per-shape grounding difficulty to human nameability (\S\ref{sec:kilogram}).

\paragraph{Transparent and training-free alternatives to CLIP.}
A line of work improves CLIP without fine-tuning by adding interpretable structure. Classification-by-description \citep{menon2023visual} scores images against LLM-generated attribute descriptions rather than bare class names, while ComCLIP \citep{jiang2024comclip} and GCLIP \citep{vongala2025compositionalimagetextmatchingretrieval} decompose a caption into entities and relations and ground each against detected sub-images. These methods share our motivation of a transparent, inspectable scoring channel, but they presuppose recognizable objects or nameable attributes to decompose and localize, precisely what an abstract tangram silhouette lacks, since it has no detectable entities and its whole-shape reading (``a dancer'') is not a composition of localizable parts. Our retrieval-plus-classical-similarity channel is a transparent alternative that does not depend on object detection. Relatedly, robustness studies such as the Perceptual Observatory \citep{anvekar2025perceptual} report that CLIP-style models degrade sharply on stylized, out-of-distribution inputs, consistent with the near-chance behavior we observe on silhouettes far from CLIP's natural-image training distribution.

\paragraph{Classical shape matching and content-based retrieval.}
The matching stage of TRACE draws on a long line of shape- and template-based recognition that predates learned embeddings: shape-context descriptors \citep{belongie2002shape}, moment invariants \citep{hu1962moment}, and the broader content-based image retrieval tradition \citep{smeulders2000cbir}, all of which compare geometry directly rather than through a learned representation. Two things distinguish our setting. First, the query is \emph{linguistic}: rather than a shape to match against a shape database, we have an utterance, which we turn into visual evidence by retrieval. Second, the match is cross-modal, a natural photograph against an abstract binary silhouette, so descriptors tuned to same-domain shape comparison do not transfer directly, and SIFT alignment plus a structural quality index is one combination that bridges the photo--silhouette gap (\S\ref{sec:pipeline}). A systematic comparison of our scorer against classical shape descriptors (shape context, Hu or Zernike moments) is a natural next step we leave to future work.

\paragraph{Rational Speech Acts and adaptive generation.}
The dominant computational-pragmatics paradigm for reference is Rational Speech Acts \citep{frank2012rsa,goodman2016}, with partner-adaptation extensions \citep{zhu2021}. RSA is speaker-centric, specifying how a speaker chooses utterances for a modeled listener, and complementary to our baseline, which recovers a candidate set a downstream RSA module could refine.

\paragraph{Emergent communication and language evolution.}
A substantial literature studies how conventions form in iterated multi-agent settings, in the Steels naming-game tradition \citep{steels2012naming,beuls2013} and in deep-learning emergent communication \citep{lazaridou2020,rita2025}. These study how conventions arise; we do not. An inspectable listener-side pact state (future work, \S\ref{sec:discussion}) could plug into such a system without changing any other component.

\paragraph{Multimodal convention formation.}
\citet{maeda2026} extend probabilistic convention-formation models to include gesture, and document that human partners shift modality preferences across repetitions. Their findings, along with~\cite{bingham2026a,bingham2026multimodalframeworkaligninghuman}, support a modality-agnostic view of listener-side pact state, the future direction we sketch in \S\ref{sec:discussion}.

\paragraph{Explicit symbolic state representations.}
The future direction we sketch (\S\ref{sec:discussion}), an explicit, inspectable representation of listener-side pact state, aligns with a tradition in dialogue state tracking (DST) of maintaining explicit state rather than an opaque neural memory. DSGFNet \citep{feng2022dsgfnet}, for example, evolves a dynamic schema graph across turns to fuse prior slot--domain membership with dialogue-aware slot relations. Where DST tracks task-slot values on the system side, such a representation would track referent--object pacts on the listener side; the shared commitment is to a state that accumulates across turns in a form open to inspection.

\section{Task and Data}
\label{sec:task}

We use the cued variant of the Stanford Repeated Reference Game corpus \citep{hawkins2020,hawkins2019arxiv}, which records 7,867 director--matcher messages across 83 games of 12 tangram figures (labeled A through L). In each of six repetition blocks, the director sees a private target cue and produces one or more messages; the matcher clicks a tangram; feedback is given after each trial. We use only director messages; matcher messages, choices, and timing are withheld from all models.

For each experiment we extract the \emph{first director utterance per trial}, which corresponds to the setting of the human baseline reported by \citet{hawkins2020}: matchers in the ``silent'' condition committed to a tangram based on the director's first message alone. Two subsets: (i) \emph{round-1-only}, restricting to first-repetition trials (n=991), which is the cleanest analog to the human baseline reported in \citet{hawkins2020} page 8; and (ii) \emph{all-rounds}, including all 5,969 first-utterance-per-round trials. Chance accuracy is 1/12 = 8.33\%.

\paragraph{Human baseline.}
On first-repetition trials, \citet{hawkins2020} report a 23\% error rate for matchers committing without backchannel dialogue (77\% accuracy) and a 20\% error rate for matchers who did engage in dialogue (80\% accuracy); the difference is not significant. \citet{zhao2025} independently report 78\% round-1 human accuracy on the same corpus. These are the human baselines we compare against below.

\section{CLIP as an End-to-End Baseline}
\label{sec:clip}

We first establish what a modern vision--language embedding achieves on this task without any pact-tracking or retrieval infrastructure. Using \texttt{openai/clip-vit-base-patch32} without fine-tuning, we compute the CLIP text embedding of each director utterance and the CLIP image embedding of each of the 12 tangrams, L2-normalize both, and rank tangrams by cosine similarity to the utterance. We report top-$k$ accuracy with percentile bootstrap 95\% confidence intervals (2000 resamples).

CLIP is well above chance at every $k$ (17.76\% top-1, 39.86\% top-3, and 55.30\% top-5), indicating it does extract some grounding signal from utterance to tangram. But it is well below the human baseline: top-1 at 17.76\% is less than a quarter of the $\sim$77\% humans achieve on the same trials, and top-5 at 55.30\% still leaves the correct target outside CLIP's top-5 hypothesis roughly 45\% of the time.

Three properties of the task make this outcome expected in retrospect. First, tangrams are abstract black silhouettes, adversarial to CLIP's training distribution of natural photographs. Second, the human utterances are informal and often metaphorical (``spy vs. spy guy walking with a flag,'' ``lady facing right diamond head''), differing from the descriptive captions CLIP was trained on. Third, the 12 tangrams share substantial visual structure (all are compositions of the same 7 pieces), so even when CLIP correctly identifies a general category like ``human-like silhouette'' it cannot easily discriminate between silhouettes within that category.

The CLIP setup above is deliberately minimal: CLIP's smallest standard backbone, no domain cue, and no augmentation, whereas the retrieval channel of \S\ref{sec:main} receives a ``tangram figure'' cue and rotational augmentation. \S\ref{sec:fair} tests whether giving CLIP those advantages closes the gap.

\subsection{Prompt, augmentation, and scale controls}
\label{sec:fair}

To test whether this understates what an embedding scorer can do here, we vary three factors the retrieval channel benefits from: a domain-anchoring prompt, orientation augmentation, and model scale. For the prompt condition we embed each utterance through an ensemble of four templates that name the domain explicitly (e.g.\ ``a black tangram silhouette of \{utterance\}'') and average the normalized text embeddings. For the orientation condition we score each tangram at its best-matching rotation (max cosine over $\{0, 90, 180, 270\}$ degrees). We cross both factors and run the full $2\times2$ grid on two backbones: the base \texttt{ViT-B/32} above and the larger \texttt{ViT-L/14} \citep{radford2021clip}.

\begin{table}[t]
\centering
\small
\setlength{\tabcolsep}{4pt}
\begin{tabular}{llccc}
\toprule
Backbone & Condition & top-1 & top-3 & top-5 \\
\midrule
\texttt{ViT-B/32} & raw              & 17.76 & 39.86 & 55.30 \\
                  & +prompt          & 12.41 & 34.01 & 51.46 \\
                  & +rotation        & 15.24 & 35.12 & 50.66 \\
                  & +prompt+rot.     & 11.10 & 30.68 & 49.45 \\
\midrule
\texttt{ViT-L/14} & raw              & \textbf{26.24} & \textbf{44.40} & \textbf{59.94} \\
                  & +prompt          & 20.28 & 39.15 & 58.02 \\
                  & +rotation        & 22.30 & 41.27 & 57.32 \\
                  & +prompt+rot.     & 18.77 & 36.53 & 51.36 \\
\bottomrule
\end{tabular}
\caption{Prompt, augmentation, and scale controls, round-1 top-$k$ (\%, n=991). Neither the domain prompt nor best-of-rotations improves either backbone; both reduce accuracy. Model scale is the only factor that helps: \texttt{ViT-L/14} raw is the best CLIP configuration and is the number carried into Table~\ref{tab:main}. Chance is 8.33\%.}
\label{tab:fair}
\end{table}

Results are in Table~\ref{tab:fair}. Two findings stand out. First, neither the domain prompt nor the orientation augmentation helps; both reduce accuracy, on both backbones. The domain prompt costs 5--6 top-1 points and best-of-rotations costs 2--4. We attribute the prompt effect to variance collapse: wrapping already-descriptive utterances in a generic ``black silhouette'' frame pulls all twelve text embeddings toward a common region and shrinks the between-tangram separation the ranking depends on. Best-of-rotations hurts because the maximum over orientations is taken for all twelve candidates, not only the target, so on stimuli where CLIP's matching is already weak it raises distractor scores at least as much as target scores. This is the mirror image of the retrieval channel, where the same augmentation \emph{helps} (\S\ref{sec:pipeline}): augmentation aids a geometric matcher operating on shape while destabilizing a semantic embedding operating on caption-like meaning.

Second, model scale does help. The larger \texttt{ViT-L/14} backbone raises raw top-1 from 17.76\% to 26.24\% and top-5 from 55.30\% to 59.94\%. We carry this stronger configuration into the comparison of \S\ref{sec:results}, representing CLIP by the best setting we found. Even so, it remains far below the human baseline.

\paragraph{Beyond CLIP: other vision--language backbones.}
The pattern above is specific to OpenAI's CLIP. To test whether weak abstract grounding is a general property of contrastive image--text pretraining, we evaluate two further off-the-shelf families on the same round-1 task (full numbers in Table~\ref{tab:main}): OpenCLIP/LAION checkpoints and SigLIP. The LAION-trained CLIP checkpoints behave like OpenAI's or slightly worse (15.0--15.4\% top-1, with no benefit from scale). SigLIP, however, is markedly stronger: 31.89\% top-1 for the base model and 39.46\% for the large one, more than double OpenAI \texttt{B/32} and above even \texttt{L/14}. Weak abstract grounding is therefore \emph{not} an intrinsic limitation of contrastive image--text models: it varies sharply across pretraining recipes, and the strongest model we tested (SigLIP-large, with higher input resolution and a sigmoid objective) approaches the classical retrieval channel (\S\ref{sec:results}). The KiloGram and SND analyses below use OpenAI CLIP and characterize that model specifically.

We observe a large, consistent gap: SigLIP grounds these silhouettes far better than any CLIP or OpenCLIP checkpoint. A plausible reason is that a tangram has no texture and no nameable parts, so grounding depends on global shape, the structural signal CLIP is documented to under-use, tending instead toward a coarse, bag-of-words alignment \citep{yuksekgonul2023bow}. SigLIP differs from the CLIP variants in training objective (a pairwise sigmoid loss rather than softmax contrastive) \citep{zhai2023siglip}, in spatial granularity (patch-16 rather than patch-32), and in pretraining data; any of these could contribute, and our experiments do not separate them. We therefore report the effect as a diagnostic observation (abstract-grounding ability varies sharply with pretraining recipe) and leave its causal decomposition to future work.

\subsection{Generalization to KiloGram}
\label{sec:kilogram}

To test whether the CLIP result of \S\ref{sec:clip} is specific to the twelve Hawkins tangrams, we apply the same scorer to KiloGram \citep{ji2022kilogram}: 1{,}013 tangram silhouettes with human whole-shape descriptions (10{,}173 description--shape trials). We evaluate two conditions. \emph{Full} ranks each description's target among all 1{,}013 shapes (chance 0.10\% top-1); \emph{matched} ranks the target within a random 12-shape context (chance 8.33\%), directly comparable to the Hawkins setup. We correlate per-shape grounding quality (mean reciprocal rank, MRR) with each shape's SND.

\begin{table}[t]
\centering
\small
\begin{tabular}{llcc}
\toprule
Condition & Metric & \texttt{B/32} & \texttt{L/14} \\
\midrule
Full (1{,}013-way) & top-1 & 1.09 & 1.68 \\
                   & top-5 & 3.35 & 6.04 \\
                   & top-10 & 5.54 & 8.99 \\
\midrule
Matched (12-way)   & top-1 & 20.58 & 25.27 \\
                   & top-5 & 60.62 & 64.45 \\
\midrule
SND corr.\ (MRR)   & $\rho$ & $-0.14$ & $-0.26$ \\
\bottomrule
\end{tabular}
\caption{CLIP on KiloGram \citep{ji2022kilogram}: 1{,}013 shapes, 10{,}173 description--shape trials. Full-ranking chance is 0.10 / 0.49 / 0.99\% at top-1/5/10; matched 12-way chance is 8.33\%. The matched accuracies track the Hawkins CLIP numbers (17.76\% \texttt{B/32}, 26.24\% \texttt{L/14}; \S\ref{sec:clip}). SND correlation is Spearman $\rho$ between per-shape MRR and Shape Naming Divergence; both $p<10^{-5}$.}
\label{tab:kilogram}
\end{table}

Results are in Table~\ref{tab:kilogram}. Three findings. First, on the full 1{,}013-way ranking CLIP is far above chance but low in absolute terms (1.09\% / 1.68\% top-1 for \texttt{ViT-B/32} / \texttt{ViT-L/14}), confirming that abstract-silhouette grounding is hard well beyond the twelve Hawkins figures. Second, and more informatively, the matched 12-way accuracies (20.58\% and 25.27\%) closely track the Hawkins numbers of \S\ref{sec:clip} (17.76\% and 26.24\%): CLIP's abstract-grounding accuracy is stable across two independent tangram sets that differ in size by nearly two orders of magnitude. Third, per-shape grounding quality correlates negatively and significantly with SND (Spearman $\rho = -0.14$, $p < 10^{-5}$ for \texttt{ViT-B/32}; $\rho = -0.26$, $p < 10^{-16}$ for \texttt{ViT-L/14}): CLIP grounds the shapes humans find harder to name less well. The effect is stronger for the more capable backbone, so as CLIP improves its residual errors concentrate on precisely the shapes that are hard for humans too, a convergence of model and human difficulty rather than a divergence. The correlation is modest in magnitude ($\rho=-0.26$ accounts for roughly 7\% of per-shape variance); we read it as a consistent trend, not a dominant factor. SND is KiloGram's own nameability norm; we correlate against it rather than recomputing it.

\section{A Transparent Retrieval Baseline}
\label{sec:main}

We now describe the transparent retrieval baseline: a retrieval-based perceptual channel scored by a classical shape-similarity metric, with no learned visual representation.

\begin{figure*}[t]
\centering
\includegraphics[width=\textwidth]{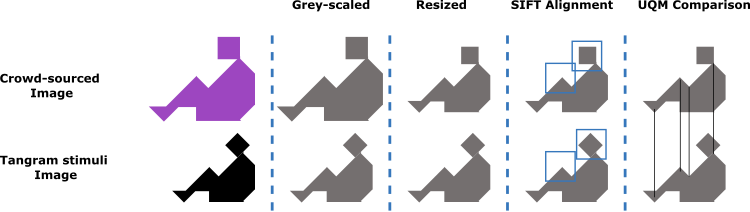}
\caption{The transparent scoring pipeline. Each crowd-sourced retrieved image and each tangram stimulus is grey-scaled, resized, and aligned to the tangram with a SIFT homography; the Universal Quality Index is then computed on the aligned pair. The tangram's score $g(o_i, I_\varphi)$ is the best such value over the retrieved set and its rotational and grayscale-inverted augmentations.}
\label{fig:pipeline}
\end{figure*}

\subsection{Retrieval-based perceptual channel}
\label{sec:pipeline}

We call this transparent pipeline TRACE. For each director utterance $\varphi$, we (i) normalize the utterance with spaCy \citep{spacy}, retaining only tokens whose part-of-speech tag is noun, verb, or conjunction; (ii) append the domain cue ``tangram figure''; (iii) submit the transformed query $q_\varphi$ to a commercial image search API \citep{bing} and retain the top $k=7$ retrievals as $I_\varphi$; and (iv) for each tangram $o_i$, align $I_\varphi$ to $o_i$ via SIFT homographies \citep{lowe2004sift,lindeberg2012}, apply rotational and grayscale-inverted augmentation, and compute a similarity score $g(o_i, I_\varphi)$ using the Universal Quality Index \citep{wang2002uqi}. Given a fixed retrieved set $I_\varphi$, the pipeline is deterministic: the same images yield the same alignments, similarity scores, and ranking on every run, so the reported accuracies require no averaging over repeated retrieval passes; there is no run-to-run variation to average out. The preprocessing choices are deliberate: retaining nouns, verbs, and conjunctions keeps the content and relational words that describe a figure while dropping function words that add retrieval noise; the ``tangram figure'' cue steers the search toward silhouette-like results rather than unrelated senses of a word; and $k=7$ maximizes accuracy over a broad $k=5$--$7$ plateau (Figure~\ref{fig:accnum}), so the setting is not finely tuned. The augmentation targets nuisance variation in retrieved photographs: they differ in orientation and in figure--ground polarity (a dark figure on a light background, or the reverse), so scoring each tangram against rotated and grayscale-inverted copies lets a correct match survive these variations rather than be lost to a single fixed alignment.

We deliberately use classical similarity rather than a learned embedding as the similarity function for two reasons. First, we wanted to keep the perceptual channel inspectable end to end. Second, and connected: the CLIP baseline of \S\ref{sec:clip} demonstrates that direct learned similarity underperforms on this task in isolation.

We test directly whether a \emph{learned} similarity would do as well \emph{inside} the retrieval channel: holding retrieval fixed, we replace SIFT+UQI with the cosine between the retrieved images and each tangram in a learned embedding space. Both OpenAI CLIP and SigLIP-large, the strongest end-to-end VLM in our comparison, collapse to near chance on this variant (9.1\% top-1), far below their own end-to-end grounding and far below SIFT+UQI's 44.3\%. Giving a learned model the same retrieved images therefore does not help: matching a natural photograph to an abstract silhouette in a learned embedding space is uninformative, whereas the classical shape alignment of SIFT+UQI bridges that gap. This isolates the source of the pipeline's advantage, a shape-appropriate classical similarity rather than privileged access to external images, and speaks directly to the fairness of the comparison: the strongest VLM, handed the very images the retrieval channel uses, cannot exploit them.

\begin{figure*}[t]
\centering
\includegraphics[width=\textwidth]{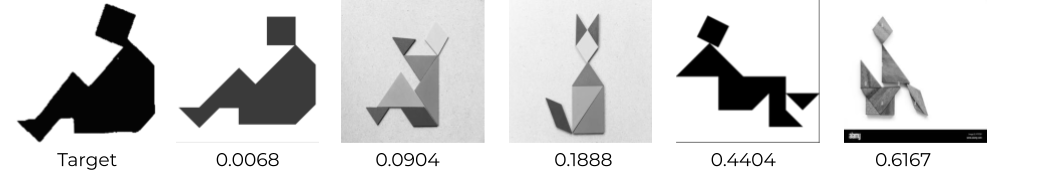}
\caption{UQI as a graded similarity signal. A target tangram (left) scored against five candidate images, with the UQI value shown beneath each (here ranging from $0.007$ to $0.617$). UQI yields a smooth, shape-sensitive score rather than a binary match, and the pipeline ranks the twelve tangrams by the best such score over the retrieved set.}
\label{fig:continuum}
\end{figure*}

The choice of similarity metric matters (Table~\ref{tab:metrics}, Appendix~\ref{app:ablations}; Figure~\ref{fig:continuum} shows UQI grading a target against candidate matches). The 15.8-point gap between the two structural metrics has a simple cause: UQI \citep{wang2002uqi} is the $C_1=C_2=0$ special case of SSIM \citep{wang2004ssim}, whose stabilizing constants push the large uniform-background regions of a binarized silhouette toward a similarity of $1$ regardless of shape, compressing the discriminative range that UQI, dominated instead by the high-variance boundary, retains. Augmentation (rotational copies plus grayscale inversion) adds 8.4 points on the same ablation (33.3\% to 41.66\%).

\subsection{Main results}
\label{sec:results}

Table~\ref{tab:main} reports single-utterance top-$k$ accuracy for the retrieval channel and six off-the-shelf vision--language backbones (OpenAI CLIP, OpenCLIP/LAION, and SigLIP) on identical trials, with bootstrap 95\% confidence intervals.

\begin{table*}[t]
\centering
\small
\begin{tabular}{lccc}
\toprule
Method & top-1 & top-3 & top-5 \\
\midrule
TRACE (this work) & \textbf{44.3} {\footnotesize[37, 52]} & \textbf{65.2} {\footnotesize[58, 73]} & \textbf{82.2} {\footnotesize[76, 88]} \\
SigLIP-large & 43.0 {\footnotesize[36, 50]} & 61.8 {\footnotesize[55, 69]} & 76.4 {\footnotesize[70, 82]} \\
SigLIP-base & 33.3 {\footnotesize[26, 40]} & 58.8 {\footnotesize[52, 66]} & 70.3 {\footnotesize[64, 77]} \\
CLIP \texttt{L/14} (OpenAI) & 26.7 {\footnotesize[20, 33]} & 41.2 {\footnotesize[34, 48]} & 60.0 {\footnotesize[53, 67]} \\
CLIP \texttt{B/32} (OpenAI) & 18.8 {\footnotesize[13, 25]} & 40.6 {\footnotesize[33, 48]} & 58.8 {\footnotesize[52, 66]} \\
OpenCLIP \texttt{B/32} (LAION) & 17.6 {\footnotesize[12, 23]} & 39.4 {\footnotesize[32, 47]} & 53.3 {\footnotesize[46, 61]} \\
OpenCLIP \texttt{L/14} (LAION) & 12.7 {\footnotesize[8, 18]} & 32.7 {\footnotesize[25, 39]} & 52.7 {\footnotesize[45, 60]} \\
\bottomrule
\end{tabular}
\caption{Single-utterance top-$k$ accuracy on the data ($n=991$: the first-repetition trials for which the retrieval channel returned images, so that every method is scored on identical trials), in \% with bootstrap 95\% CIs. Chance is 8.33\%; rows ordered by top-1. The transparent retrieval channel is statistically indistinguishable from SigLIP-large at every $k$ (overlapping intervals), and the retrieval and SigLIP-large top-1 intervals separate from every CLIP/OpenCLIP variant. Human top-1 on the corpus is $\approx$77--80\% \citep{hawkins2020}.}
\label{tab:main}
\end{table*}

On these identical trials, the transparent retrieval channel and SigLIP-large are statistically indistinguishable at every $k$: their 95\% intervals overlap at top-1 (44.3 [37, 52] vs.\ 43.0 [36, 50]), top-3, and top-5, with the retrieval channel's point estimates marginally higher. Both, in turn, separate from the remaining models: the retrieval and SigLIP-large top-1 intervals do not overlap CLIP-\texttt{L/14}'s [20, 33], and the gap to OpenCLIP is larger still. Because every model is scored on the identical trials, we complement the intervals with paired tests against each model (Table~\ref{tab:paired}).

\begin{table}[t]
\centering
\small
\begin{tabular}{lcccc}
\toprule
vs Model & acc & diff & 95\% CI & $p$ \\
\midrule
SigLIP-large           & 43.0\% & +1.3  & [$-8$, +11]  & 0.82 \\
SigLIP-base            & 33.3\% & +11.0 & [+2, +20]    & 0.02 \\
CLIP \texttt{L/14}     & 26.7\% & +17.6 & [+9, +26]    & $e-4$ \\
CLIP \texttt{B/32}     & 18.8\% & +25.5 & [+17, +34]   & $e{-8}$ \\
OpenCLIP \texttt{B/32} & 17.6\% & +26.7 & [+18, +35]   & $e-9$ \\
OpenCLIP \texttt{L/14} & 12.7\% & +31.6 & [+24, +40]   & $e-12$ \\
\bottomrule
\end{tabular}
\caption{Paired comparison of the transparent retrieval channel (TRACE) against each vision--language model on the data. For each model: its top-1 accuracy (\%), the top-1 accuracy difference from TRACE (TRACE $-$ model, in points), the paired bootstrap 95\% confidence interval of that difference (2{,}000 resamples over trials), and the two-sided exact McNemar $p$-value on the per-trial hit vectors. Positive differences favor TRACE.}
\label{tab:paired}
\end{table}

A transparent pipeline with \emph{no learned visual representation}, classical keypoint matching and a signal-quality index, matches the strongest off-the-shelf VLM on this task and outperforms the other five, on identical trials with confidence intervals. The dataset is the 991 first-repetition trials for which the retrieval channel returned images, scored identically for every method. All systems consume the same director utterance and rank the same twelve tangrams, so the cross-system comparison is the informative one; the human baseline (77--80\%) draws on interactive dialogue and world knowledge these systems do not have.

\subsection{Where each system succeeds and fails}
\label{sec:error-analysis}

Manual inspection of first-repetition trials reveals three recognizable failure modes for the retrieval pipeline, all of which correspond to utterances where perceptual bootstrapping through web imagery is either uninformative or misleading:
\begin{enumerate}
    \item \emph{Purely geometric descriptions} (``zig zag with square on top,'' ``triangle shape on the top''). These treat the tangram as a 2D pattern rather than a depicted object; retrieval returns generic solved-square tangrams that match every target roughly equally.
    \item \emph{Metaphorical or idiomatic descriptions} (``guy doing a weird dance move''). Easy for a human to visualize but hard to translate into a query whose top results resemble a silhouette.
    \item \emph{Already-entrained shorthand} (``the sitting one'' in a late round). These are pronouns-with-respect-to-a-pact that depend on the pact state established in earlier turns; the retrieval channel treats each utterance as standalone.
\end{enumerate}
CLIP's failure mode is different: it is largely uniform. Because CLIP's text and image encoders were trained on natural photography with descriptive captions, its similarity scores over the 12 tangram silhouettes show relatively little differentiation between tangrams for most utterances (mean pairwise cosine gap of $\sim$0.05). This is consistent with CLIP's round-1 pattern (Table~\ref{tab:main}): CLIP's top-5 accuracy is proportionally much higher than its top-1 (roughly 3$\times$), which is what one would expect from a system that ranks tangrams noisily around a similar mean rather than one that confidently commits to the wrong tangram. The retrieval channel, by contrast, commits more confidently and is therefore more wrong when it is wrong.

\section{Internal Validity}
\label{sec:validity}

A natural concern for a web-retrieval channel evaluated on publicly available stimuli is that the image search might surface the target tangram itself, or a near-duplicate, within the retrieved set $I_\varphi$; the UQI similarity for the correct $o_i$ would then be inflated by self-similarity rather than by grounding. To rule this out, we manually inspected the retrieved image sets for the first-repetition trials and found no cases in which the target tangram or a near-duplicate appeared among the retrieved images. The reported accuracies therefore reflect cross-modal grounding rather than retrieval of the target itself.

\section{Discussion}
\label{sec:discussion}

The overall ordering is robust: on this task, CLIP alone $<$ classical retrieval $<$ humans, and this holds for the fairest CLIP configuration (\texttt{ViT-L/14}, no prompt, no rotation). The gap between retrieval and humans is what remains for future work, and the controls (\S\ref{sec:fair}) show that closing it is unlikely to come from prompting or orientation tricks on an off-the-shelf embedding.

\paragraph{Extensions.}
These experiments speak to the single-shot setting; the repeated game that motivates them is interactive, and LVLMs are documented to lose pact state across turns \citep{imai2025,zeng2026,zhao2025}. Extending an inspectable listener-side representation of the pact state to that setting, and evaluating it with live partners, is the natural next step, which we leave to future work.

\section{Conclusion}

We asked whether grounding abstract references requires a large pretrained vision--language model. On single-utterance grounding over the twelve Stanford tangrams, the answer, for this task, is no: on identical trials with confidence intervals, a classical pipeline with \emph{no learned visual representation}, SIFT keypoint matching and a signal-quality index over retrieved images, matches the strongest off-the-shelf VLM and outperforms the other five. This is not a matched-information comparison: the baseline additionally retrieves external images through a commercial search engine, itself a learned, language-driven system. The result therefore does not show that VLMs are unnecessary; it shows that a learned \emph{visual} representation is not the bottleneck for this task, since given retrieved images a shape-appropriate classical similarity is enough.

The diagnostic behind this result is itself informative. Abstract-grounding ability varies sharply and specifically across off-the-shelf models rather than being an intrinsic limit of contrastive image--text pretraining: SigLIP grounds these silhouettes far better than any CLIP or OpenCLIP checkpoint, a gap that plausibly reflects differences in training objective, spatial granularity, or pretraining data rather than model scale alone. For OpenAI CLIP specifically, the weakness generalizes to the 1{,}013-shape KiloGram benchmark \citep{ji2022kilogram}, where per-shape grounding difficulty tracks human shape-nameability (\S\ref{sec:kilogram}).

Interactive evaluation with live partners is a natural next step, but the finding stands on its own: grounding abstract references turns on choosing a representation suited to the perceptual structure of the task, more than on the raw power of a learned one.

\section*{Limitations}
\label{sec:limitations}

\textbf{Passive listener.} We evaluate only on prerecorded director utterances; the listener never asks clarifying questions or produces confirmatory utterances. This is a fundamentally simpler task than the interactive game humans play, and the results should be read accordingly. Our contribution is a component of what a full interactive agent would need, not a full interactive agent.


\textbf{What the transparent baseline's advantage isolates.} TRACE has access to web-retrieved images that the end-to-end VLMs do not, so its advantage could in principle stem from the classical similarity, from the retrieval step, from the external image knowledge, or from their combination. Two controls narrow this down: the retrieval-source ablation (Appendix~\ref{app:ablations}, Table~\ref{tab:source}) shows the accuracy depends on \emph{utterance-relevant} retrieved content (shuffled or random images collapse it to chance), and the in-retrieval similarity control (\S\ref{sec:pipeline}) shows a learned similarity over the same images does not help. Fully isolating the marginal contribution of the retrieval step from that of the similarity function is left to future work. We note, too, that the asymmetry is one of \emph{kind} rather than sheer amount: the visual evidence TRACE consumes per decision is a handful of inspectable images (at most $k=7$) retrieved for one query, and the matching stage has no learned visual parameters, whereas each VLM brings a large image corpus compressed into its weights. This does not neutralize the retrieval advantage, since those images are freshly retrieved and utterance-specific, but it clarifies what ``access to external images'' does and does not mean here.

\textbf{Scope of the claim.} TRACE is not evidence that classical vision is generally preferable to learned representations; it shows that for this narrowly defined grounding task, transparent retrieval-based matching remains surprisingly competitive.

\textbf{Computational cost.} Transparency is not efficiency. Scoring a trial aligns each of $k$ retrieved images to all twelve tangrams under several augmentations with SIFT, which is not obviously cheaper than a single VLM forward pass; our claim is that the task does not require a \emph{learned} visual representation, not that the classical pipeline is faster.

\textbf{Reliance on a commercial search index.} The pipeline depends on a proprietary image-search API, which raises reproducibility concerns (the index and its ranking drift over time and geography), provenance and consent concerns for the returned images, and equity concerns, since paid API access is itself a barrier to reproduction. A public, versioned retrieval corpus would mitigate all three.

\textbf{English- and culture-specific.} Both the spaCy preprocessing and the ``tangram figure'' domain keyword are English-specific. The retrieval channel also inherits any cultural biases of Bing's image-search index. Multilingual and cross-cultural portability is untested and likely non-trivial.

\bibliography{custom}

\appendix

\section{Ablations}
\label{app:ablations}

\begin{figure}[t]
\centering
\includegraphics[width=\columnwidth]{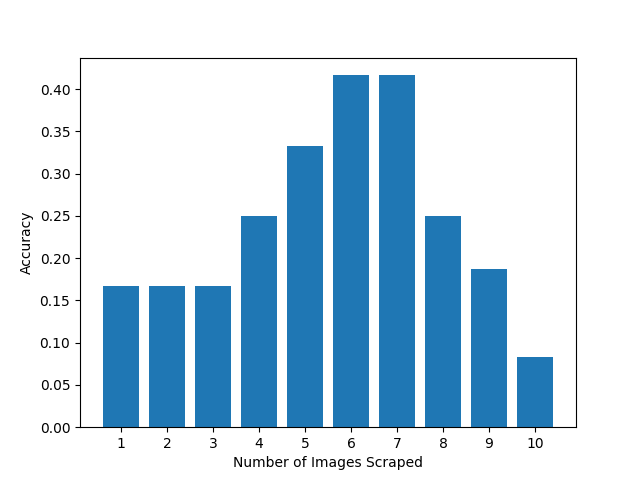}
\caption{Effect of retrieval breadth. Top-1 accuracy as a function of the number of images scraped per query. Accuracy peaks at six to seven images and declines as additional, less relevant images dilute the retrieved set; the pipeline uses $k=7$.}
\label{fig:accnum}
\end{figure}

We report three ablations for the retrieval channel. First, a retrieval-source ablation (Table~\ref{tab:source}) locates where the pipeline's accuracy comes from, on the dataset with bootstrap 95\% confidence intervals: holding the SIFT+UQI scorer fixed, we vary only which images each trial is scored against. Scoring each trial against another trial's retrievals (\emph{shuffled}) drops top-1 from 44.3\% to 6.1\% (below chance, as the scorer then confidently commits to the wrong tangram), and a fixed random pool leaves it at chance (8.5\%). The accuracy therefore depends specifically on \emph{utterance-relevant} retrieved content: neither a per-tangram bias in the scorer (which would survive shuffling) nor generic image structure (which would survive the random pool) accounts for it. Together with the in-retrieval similarity control of \S\ref{sec:pipeline} (a learned similarity over the same images does not help), this locates the accuracy in a shape-appropriate classical similarity applied to utterance-relevant retrievals. The remaining two ablations are config-selection sweeps on the full first-repetition corpus, as single deterministic passes without confidence intervals (the retrieval endpoint is now rate-limited; the Limitations section): Table~\ref{tab:metrics} compares ten classical similarity metrics under identical SIFT pre-alignment, and Figure~\ref{fig:accnum} varies the number of images retrieved per query.

\begin{table}[t]
\centering
\setlength{\tabcolsep}{4pt}
\resizebox{\columnwidth}{!}{%
\begin{tabular}{lccc}
\toprule
Retrieved images & top-1 & top-3 & top-5 \\
\midrule
Real & \textbf{44.3}\,{\footnotesize[37,52]} & \textbf{65.2}\,{\footnotesize[58,74]} & \textbf{82.2}\,{\footnotesize[77,88]} \\
Shuffled & 6.1\,{\footnotesize[2,10]} & 23.6\,{\footnotesize[17,30]} & 38.8\,{\footnotesize[31,47]} \\
Random & 8.5\,{\footnotesize[4,13]} & 25.5\,{\footnotesize[19,32]} & 41.2\,{\footnotesize[33,49]} \\
\bottomrule
\end{tabular}%
}
\caption{Retrieval-source ablation on the data, holding the SIFT+UQI scorer fixed and varying only which images each trial is scored against. \emph{Real}: the trial's own utterance-conditioned retrievals. \emph{Shuffled}: another trial's retrievals (a random derangement). \emph{Random}: one fixed random sample reused for every trial. Bootstrap 95\% CIs (rounded); chance is 8.33\%.}
\label{tab:source}
\end{table}

\begin{table}[t]
\centering
\small
\begin{tabular}{lc}
\toprule
Metric (with SIFT alignment) & Top-1 \\
\midrule
Universal Quality Index (UQI) & \textbf{41.66\%} \\
Structural Similarity (SSIM) & 25.8\% \\
Peak Signal-to-Noise (PSNR) & 19.7\% \\
Mean Squared Error & 18.2\% \\
Mean Absolute Error & 17.5\% \\
ERGAS & 16.8\% \\
Spatial Correlation Coefficient & 15.4\% \\
RASE & 14.9\% \\
Spectral Angle Mapper & 14.2\% \\
Visual Information Fidelity & 13.7\% \\
\bottomrule
\end{tabular}
\caption{Ten classical similarity metrics evaluated with SIFT pre-alignment as a config-selection ablation. UQI leads the runner-up (SSIM) by 15.8 points. These are single-pass point estimates without confidence intervals (the retrieval endpoint is now rate-limited; the Limitations section). Chance is 8.33\%.}
\label{tab:metrics}
\end{table}

\section{Evaluation Protocol}
\label{app:protocol}

For reproducibility we specify exactly how each method receives its inputs. All six vision--language models are off-the-shelf \emph{contrastive dual-encoders} scored by cosine similarity; none is prompted generatively or shown in-context examples. Every method is evaluated on the same trials and the same twelve tangram silhouettes (\texttt{tangram\_A.png}--\texttt{tangram\_L.png}).

\paragraph{Vision--language models (main comparison).}
\emph{Text input:} the raw director utterance, fed verbatim to the model's text encoder through its native processor (with the processor's default tokenization, padding, and truncation). No prompt template, instruction, or few-shot context is added in the main condition. \emph{Image input:} each of the twelve tangram silhouettes is loaded as RGB and passed through the model's native image processor at that checkpoint's default resolution (224\,px for the four CLIP checkpoints and SigLIP-base, 256\,px for SigLIP-large), which applies the checkpoint's own resize and normalization. \emph{Scoring:} we L2-normalize the utterance text embedding and each tangram image embedding and rank the twelve tangrams by cosine similarity to the utterance; the top-1 prediction is the arg-max, and top-$k$ counts the target in the $k$ highest-scoring tangrams. For CLIP and OpenCLIP the projected embeddings are used (\texttt{text\_projection} / \texttt{visual\_projection}); for SigLIP the pooled encoder outputs. Chance is $1/12 = 8.33\%$. \emph{Checkpoints} (Hugging Face identifiers): \texttt{openai/clip-vit-base-patch32}, \texttt{openai/clip-vit-large-patch14}, \texttt{laion/CLIP-ViT-B-32-laion2B-s34B-b79K}, \texttt{laion/CLIP-ViT-L-14-laion2B-s32B-b82K}, \texttt{google/siglip-base-patch16-224}, and \texttt{google/siglip-large-patch16-256}.

\paragraph{Baseline controls (Table~\ref{tab:fair}).}
Two variants test whether the minimal CLIP setup understates its ability. \emph{Domain prompt:} instead of the bare utterance, the text encoder receives an \emph{ensemble} of tangram-referring templates (the set \texttt{DOMAIN\_TEMPLATES} in the released code; the bare-utterance template is deliberately excluded), with the per-template text embeddings averaged and then L2-renormalized, following standard CLIP zero-shot prompt ensembling. \emph{Best-of-rotations:} each tangram is rendered at the four cardinal rotations $\{0,90,180,270\}^\circ$ (with a white fill and \texttt{expand} so corners are not clipped) and scored by the maximum cosine over those rotations. Both \emph{reduce} accuracy (Table~\ref{tab:fair}); the main comparison therefore uses the bare utterance and unrotated tangrams, carrying the stronger \texttt{ViT-L/14} scale forward.

\paragraph{TRACE (retrieval channel).}
For each utterance we (i) parse it with spaCy and keep only noun, verb, and conjunction tokens; (ii) append the cue ``tangram figure'' to form the query $q_\varphi$; (iii) issue $q_\varphi$ to the image-search API and keep the top $k=7$ results as $I_\varphi$; and (iv) for each tangram, align every image in $I_\varphi$ to the silhouette with a SIFT homography, add rotational and grayscale-inverted copies, and score each aligned pair with UQI, taking the best value over the set. Tangrams are ranked by this score (\S\ref{sec:pipeline}). The candidate tangrams are the same twelve silhouettes used for the VLMs.

\paragraph{In-retrieval similarity control and retrieval-source ablation.}
Both reuse TRACE's retrieved image sets $I_\varphi$. The in-retrieval similarity control (\S\ref{sec:pipeline}) replaces SIFT+UQI with the cosine between each retrieved image and each tangram in a learned embedding space (CLIP or SigLIP-large, embeddings extracted exactly as above, taking the maximum over $I_\varphi$). The retrieval-source ablation (Table~\ref{tab:source}) holds SIFT+UQI fixed and substitutes $I_\varphi$ with another trial's set (\emph{shuffled}, a random derangement) or a single fixed random sample reused across trials (\emph{random pool}).

\paragraph{KiloGram (Table~\ref{tab:kilogram}).}
Each KiloGram whole-shape description is the text input and each rasterized silhouette is an image, scored by CLIP cosine as above. The \emph{full} condition ranks the target among all 1{,}013 shapes; the \emph{matched} condition ranks it within a 12-way controlled context (the target plus eleven distractors from KiloGram's controlled context batches, \texttt{eval\_batch\_data.json}), directly comparable to the twelve-tangram Hawkins setup, with mean context size $\approx 12$ and chance $\approx 8.3\%$.


\end{document}